\documentstyle[aps,epsf,multicol,pre]{revtex}
\input epsf
\begin{document}
\draft
\title{Small World Patterns in Food Webs}

\author{Jose M. Montoya$^{1,2}$ and Ricard V. Sol\'e$^{1,3}$}

\address{$^1$ Complex Systems Research Group, FEN\\
Universitat Polit\`ecnica de Catalunya, Campus Nord B4, 
08034 Barcelona, Spain\\
$^2$Intercampus Department of Ecology, University of Alcal\'a\\
28871 Alcal\'a de Henares, Madrid, Spain\\
$^3$Santa Fe Institute, 1399 Hyde Park Road, New Mexico 87501, USA}

\maketitle
\begin{abstract}

The analysis of some species-rich, well-defined food webs 
shows that they display the so called Small World behavior 
shared by a number of disparate complex systems. The three 
systems analysed (Ythan estuary web, Silwood web and the 
Little Rock lake web) have different levels of taxonomic resolution, 
but all of them involve high clustering and short path lengths 
between species. Additionally, the distribution of connections 
$P(k)$ is skewed in all the webs analysed and shows a power-law 
behavior $P(k) \propto k^{-\gamma}$ in two cases (with 
$\gamma \approx 1$). These features suggest that communities 
might be self-organized in such a way that 
high homeostasis to perturbations (with short transient 
times to recovery) would be at work. The consequences for 
ecological theory are outlined.

\end{abstract}

\begin{multicols}{2}

\section{Introduction}

The understanding of food web patterns in complex ecologies is a central 
issue in ecology. As Stuart Pimm claims, food webs are the road-maps 
through Darwin's entangled bank (Pimm et al, 1991). Beyond the specific 
features of currently described food webs (such as the concrete species 
composition) a number of regularities have been observed suggestive of 
fundamental laws of ecosystem organization (see for example Warren, 1994; 
Polis and Strong, 1996 and references therein). 

One of the most successful approaches to real food webs deals with 
the graph patterns associated with the set of links between species. 
Specifically, a food web can be described in terms of a graph $G(V,E)$ 
consisting of a finite set $V$ of {\em vertices}
(nodes, species) and a finite
set $E$ of {\em edges} such that each edge $e$ is associated with a pair 
of vertices $v$ and $w$. In spite that ecological graphs are directed (i. e. 
links go from one species to another and the reverse connection will be 
typically different in strength and sign) most theoretical studies deal 
with non-directed graphs. In other words, two species appear connected 
if they share a common edge, irrespective of its particular properties. 
This approach has been rather successful, since many reported regularities 
form field data are fully recovered from non-directed graph arguments of 
different nature (Cohen et al., 1990; Pimm et al., 1991; 
Williams and Martinez, 2000; see also Albert et al. 2000). 
This observation is probably a consequence 
of fundamental laws constraining the basic topological arrangements 
of ecological networks that go beyond the specific rules of dynamical 
interaction (Brown, 1994; Sol\'e and Bascompte, 2001). 

Recently, new theoretical approaches to graph complexity in nature have 
emerged. Two main results of these studies are: (a) the widespread presence of 
the so called {\em small world} (SW) pattern of 
some economic, technologic and biological networks (Watts and Strogatz, 1998; 
Watts, 1999; Adamic, 1999) and (b) the presence of scale-free (SF)  
distributions $P(k)$ of connections (Barab\'asi and Albert, 1999; 
Jeong et al., 2000; Wagner and Fell, 2000). Specifically, in 
some cases the number of nodes with 
$k$ links follows a power law distribution $P(k) \sim k^{-\gamma}$ where 
most units are weakly connected and a few are highly linked to other nodes. 
This is however a non-universal feature: the graph of the neural 
network of {\em C. elegans}, for
example, displays an exponential distribution of connections (Amaral et al., 
2000).

The SW pattern can be detected from the analysis of two basic statistical 
properties: the so called {\em clustering coefficient} $C_v$ and the 
{\em path length} $L$. Let us consider the set of links 
$\xi_{ij} (i, j, = 1, ..., S)$ where $S$ is the number of species and 
let's assume that $\xi_{ij}=1$ if a link exist and zero otherwise. Species 
will be labeled as $s_i (i=1,...,S)$. If a food web is considered,
an additional parameter is the community matrix connectance $C$. Let us
consider a given species (the i-th one) and the set of nearest neighbors 
$\Gamma_i=\{ s_i \vert \xi_{ij} = 1 \}$. Here we can calculate the clustering 
coefficient for this species as the number of connections between the species 
belonging to $\Gamma_i$. By defining 
$${\cal L}_i = \sum_{j=1}^S \xi_{ij} 
\left [ 
\sum_{k \in \Gamma_i} \xi_{jk} 
\right ] \eqno(1)$$
and thus $c_v(i)={\cal L}_i/(CS(CS-1)/2)$ so that the clustering coefficient is 
the average over all species: 
$$ C_v= {1 \over S} \sum_{i=1}^S c_v(i) \eqno(2)$$
and measures the average fraction of pairs of neighbors of a node 
that are also neighbors of each other. 

The second measure is easily defined. Given two arbitrary species 
$s_i$ and $s_j$, let $L_{min}(i,j)$ be the minimum path length 
connecting these two species. The average path length $L$ will be: 
$$ L = {2 \over S(S-1) } 
\sum_{i=1}^S \sum_{j=1}^S L_{min}(i,j) \eqno(3)$$

Small World graphs are highly clustered but the minimum path length 
between any two randomly chosen nodes in the graph is short. By comparison, 
random graphs (where nodes are randomly connected with some probability) 
are not clustered and have short $L$ (Watts, 1999). 
At the other extreme, regular lattices are typically clustered and 
have long distances. It has been shown, however, that a regular lattice can be 
transformed in to a SW if a small fraction of nodes are rewired to 
randomly chosen nodes. Thus a small degree of disorder makes the 
lattice to have short paths (as in the random case) but is still mostly 
regular (Watts and Strogatz, 1998).  

In a graph with an average of $<k>$ links per node, it can be shown 
for random graphs that $C_v^{rand} \approx <k>/N$
and $L^{rand}$ will be short. For
large networks, a SW is present if $L^{rand}$ is larger, but close to $L$ 
and $C_v \gg C_v^{rand}$.
When networks composed by a small number of units are analysed, the 
second condition is often replaced by $C_v > C_v^{rand}$. This is the case for 
the SW pattern in the neural network of the nematode worm 
{\em C. elegans} (ce) 
which has $282$ neurons and $C_v^{ce}=0.28$ and $L^{ce}=2.65$, to be 
compared with the corresponding random graphs: $C_v^{rand}=0.05$ and $L^{rand}=2.25$.
For the metabolic network of {\em E. coli}, Wagner and Fell obtained 
similar results, but with a much larger difference in the clustering. 

The consequences of the SW and SF paterns are far from trivial and can be 
of great importance in recognizing evolutionary paths, the origins of 
homeostatic stability and the sensibility to perturbations in biological 
networks. Watts and Strogatz discuss some of these ideas in their 
seminal work, suggesting that a SW architecture would play a 
relevant role in enhancing synchronization in the visual cortex. 
For metabolic networks, Wagner and Fell 
suggest that it might play a relevant role in allowing metabolism to react 
rapidly to perturbations thus displaying a very high homeostasis. What
about ecological networks? Do food webs display this type of topological
properties?

In this paper we explore this question by analysing
the statistical properties of a set of large ($S>100$) 
ecological networks where a fine taxonomic resolution is available. 
For these webs, the SW property is shown to be present and a skewed 
distribution of links is also shown to be involved in most cases. The 
presence of these properties will be shown to have deep implications 
for theoretical and applied issues at different levels.

\section{Ecological Small Worlds}

It is interesting to see that in fact most static graph models of 
food webs involve features ranging from purely random nets (constant 
connectivity networks, see Martinez, 1991a) to purely hierarchical 
models (the cascade model, see Cohen et al., 1990). 
Under the second class, a trophic species can only prey on a trophic species 
of lesser rank leading to a hierarchical structure sharing some features with 
some standard graph patterns such as Cayley
trees (Watts, 1999). For the constant connectance model
(Martinez, 1991a) no such ranking is introduced and a trophic species 
preys on any other trophic species (including itself). However, 
ecological nets display both types of properties and some studies 
suggest that they also display clustering. 

Evidence for clustering in food webs has been compiled in recent years 
(see Solow and Beet, 1998). A number of authors have shown that food webs 
are typically non-random but the effects such topological 
properties on stability are far from clear (see Solow et al., 1999, 
for a recent discussion). 

\begin{figure}
\leavevmode
\epsfxsize=7cm
\epsffile{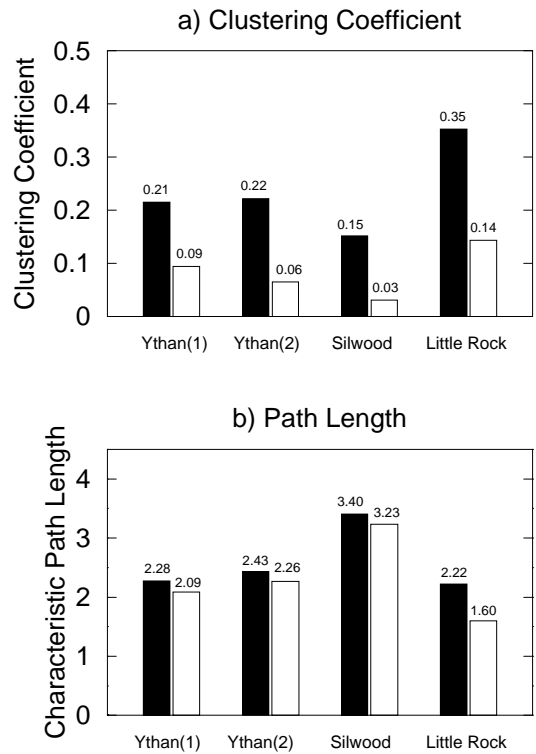}
\vspace{0.5cm}
\caption{(a) clustering coefficient for the four analysed food webs (see text). Here 
the dark bars are real webs and white bars correspond to randomly 
generated webs with the same average number of links per species (averaged 
over $200$ generated samples). We can see that in all cases the clustering 
is clearly larger than random; (b) Characteristic path length $L$ with 
random values. Except for Little Rock, the difference between the random 
and real case is very small. For the two Ythan webs, the improvement in 
resolution slightly modified the observed relations towards a better 
defined SW pattern.}
\end{figure}

The possible importance for clustering on the stability of ecological 
networks was already recognized from early statistical approaches 
(May, 1974; Pimm, 1982). Given the potential implications of the SW phenomenon for 
network homeostasis, the presence of such properties in real food webs 
would be of great interest in this discussion. 
We have found that, in fact, rich-species food webs with a good taxonomic 
resolution display the properties of small world behavior. In figure (1) 
we show the obtained results for the four large food webs analyzed, which are 
known in the literature because they are the most detailed large
community food webs available. These are: 

\begin{figure}
\leavevmode
\epsfxsize=8cm
\epsffile{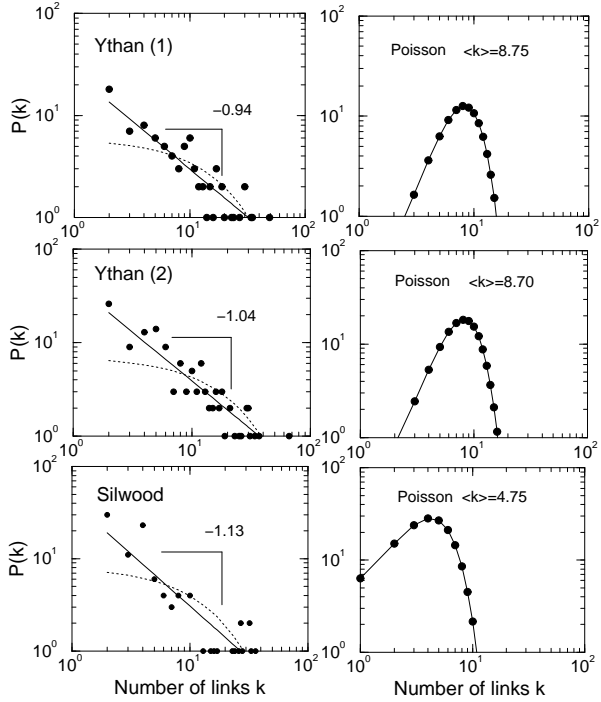}
\vspace{0.5 cm}
\caption{
Frequency distributions of links $P(k)$ for the analysed webs. Left 
column: actual data, which fit quite well a power law for the 
Ythan and Silwood cases (the exponential fit is also shown as 
a dotted line). The improvement in taxonomic resolution 
for Ythan (2) only modifies the exponent slightly. Right column:  
Expected distributions from the random graph approximation, predicting 
Poissonian behavior. None of the analysed food webs can be approached by 
the random case.}
\end{figure}

\begin{enumerate}

\item
Ythan estuary: this web is from a freshwater-marine 
interface. The number of species reported from two different studies 
are $S=93$ (Hall and Raffaelli, 1991) and $S=134$ 
(Huxham et al., 1996). These studies 
differ in the taxonomic detail incorporated and we use them in order 
to check the sensitivity in relation with the measurement of SW and scaling 
properties. This is the largest documented web in the UK and the 
average number of links per species are $<k>=8.75$ and $<k>=8.70$, 
respectively. The second web was expanded by adding 42 metazoan parasite 
species but their effect on $<k>$ was weak. For this web most ($ 88 \%$) 
nodes correspond to real species. Some nodes involve lower resolution 
at different levels (Nematodes, Acarina or Brown algae, for example, are 
lumped together).

\item 
Silwood park web (Memmott et al., 2000): this is a sub-web obtained from 
a field site of 97 hectares in size. This is a very detailed 
community centred on the Scotch broom {\em Cytisus scoparius}, involving 
$S=154$ species: one plant, 19 herbivores, 5 omnivores, 66 parasitoids, 
60 predators and three pathogens. The average connectivity is $<k>=4.75$. 
All nodes of this (sub-) network are species (except node $122$: immature spiders). 

\item 
Little Rock lake (Martinez, 1991): a web from a freshwater 
habitat, it includes $S=182$ species. The Little Rock lake is a small 
lake with an area of 18 hectares. Here $<k>=26.05$. For this web, only $31 \%$ 
of nodes are associated to species. Most are genera-level nodes ($63 \%$) and 
the rest correspond to higher taxa (Bivalvia, Hirudinea, etc). 

\end{enumerate}

We compared the observed values of $C_v$ and $L$ with 
the expected ones from the randomized webs with the same number of 
(total) links. As expected from a SW pattern, we see that all have 
very similar (and very short) distances $L$ and a clustering coeficient 
from $C_v^{Yth1}/C_v^{rand}=2.33$ to
$C_v^{Silw}/C_v^{rand}=5.0$ (numerically close to the
{\em C. elegans} data) i. e. clearly larger than expected from 
random webs. For the average path length, we found that the values 
are very close except for the Little Rock case, were notable 
differences are present.

An additional and surprising feature of the observed distributions 
of connections $P(k)$ is that they show strongly non-Poissonian  
behavior (wich would be expected from random wiring). The classic 
result by Erd\"os and R\'enyi (ER) on random graphs (see Bollob\'as,
1985) with $S$ nodes
shows that the probability that a vertex has $k$ edges follows 
a Poisson distribution 
$P(k)=e^{-<k>}\lambda^k/k!$. Here 
$$<k> = { S-1 \choose k } P_r^k (1-P_r)^{S-k-1} \eqno(2)$$
where $P_r$ is the probability that two nodes are connected. 
The observed distributions for Ythan webs and the Silwood example are 
shown in figure 2 (left column). In both webs, the distribution 
was found to be strongly skewed with a good agreement with a power law fit. 
The observed values for a regression of the log-transformed data are: 
$$\gamma^{Yth1}= 0.94 \pm 0.06 \hspace{1 cm} (r^2=0.79, p<0.01)$$ 
$$\gamma^{Yth2}= 1.04 \pm 0.05 \hspace{1 cm} (r^2=0.83, p<0.01)$$  
$$\gamma^{Silw}= 1.13 \pm 0.06 \hspace{1 cm} (r^2=0.79, p<0.01)$$

For comparison, the expected connectivity  
distributions from a random, Poissonian net 
are also shown in Figure 2 (right column).
Here the average number of links per species in each web was used.

The Little Rock web (figure 3) shows a fluctuating 
distribution of connections with a very high variance, 
but no obvious recognizable standard shape, although it is 
also skewed, with a prominent peak of 21 species (involving a 
coarse-grained cluster 
of cyanobacteria and green algae predated by pelagic cladocera, 
copepods and rotifers) with $k=26$ links. These anomalies are likely 
to due to the taxonomic resolution used (mostly at genera or higher levels).   

The exponent derived from our study is rather different from the 
estimations obtained for Internet or the metabolic graphs 
suggesting that, in spite that the SW pattern together with the 
scaling law might be common to many different types of networks,
the ecological webs belong to some different class of
complex phenomena (Sol\'e et al., 1996) to be defined in future theoretical studies.

\begin{figure}
\leavevmode
\epsfxsize=8cm
\epsffile{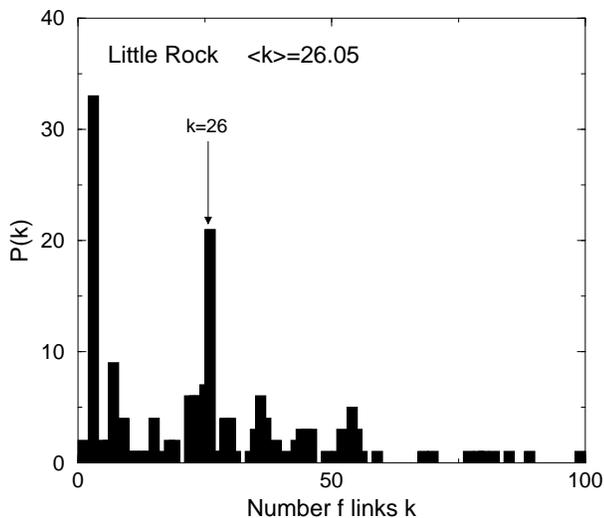}
\caption{
Little Rock lake link distribution. Here $P(k)$ is clearly skewed but 
strong fluctuations are observable as scattered peaks, in particular 
a large peak at $k=26$ involving a large cluster of interacting species 
(see text). Other peaks are indicated, revealing strong deviations from 
a monotonous power-law tail.}
\end{figure}

\section{Discussion}

In this paper we have presented evidence for SW patterns 
in ecological networks. The aim of our study was to see to what extent 
such evidence was strong enough by analysing a small number of 
large, taxonomically well-defined food webs from the ecological literature. 
The webs analysed have different origins but all of them share 
the SW pattern in their graphs and are clearly away from their 
random counterparts. As far as we know, this is the first reported 
evidence for such property in ecology. It reveals, by means of a 
new measure, the existence of clustering in food webs associated 
to short path lengths and opens a number of questions not previously 
formulated in previous theoretical studies.

Although some food web analyses found evidence for the existence
of compartments (Solow and Beet 1998), such species clusters were
established in terms of the trophic similarity between pairs of
species (see also Pimm and Lawton 1980, for a similar measure).
In contrast, the aim of the clustering coefficient as a measure
of compartmentalization is to find out to what extent groups of
species are more connected internally than they are with another
species or groups of species. In other words, this new measure go
beyond by looking not only for resource-use-guilds (see Wilson, 1999)
but also for sub-webs with higher internal connectance than the rest
of the community. 

The use of the clustering coefficient is supported by the fact that 
groups of species with
diets of similar species do not occur, especially in species-rich systems, such
as those analysed in this article (Winemiller 1990; Polis and Strong 1996). In
this respect, this measure of compartmentalization minimize the importance of
discrete and homogeneous trophic-level dynamics in complex food webs, an issue
widely considered as an artifact by many ecologists (see Polis and Strong, 
1996, for a review). The clustering coefficient does not consider the 
trophic position of the species in the food web, focusing only in the 
presence or absence of the interactions.

Two of the webs analysed correspond to the same system: the
Ythan estuary food web. These two webs involve different
taxonomic resolution but the exponents and indices remained basically 
the same for the two cases. This agreement suggests that the SW property and 
the scaling law are rather robust and that no completely detailed description 
of the species-level graph needs to be defined in order to detect these 
properties. For the best defined web (Silwood), where all nodes belong to 
species, we obtain the best evidence of SW, suggesting that less-detailed 
webs introduce a coarse-grained resolution that might hidden a higher 
degree of clustering (this seems to be confirmed by the Ythan example). 
The worst result is given by Little Rock lake, where the resolution is mainly 
at the genus level and strong deviations from the power law are observable.

The SW phenomenon appears to be associated with highly 
non-random, fat-tailed distributions of links. The plots of $P(k)$ show 
that all food webs (and others not explored here due to their 
lower taxonomic resolution) involve strong deviations from the 
Poissonian behavior and confirm the presence of clustering 
in ecological interactions. These properties reveal that some new 
features of ecological graphs have to be considered by future
models in order to explain their origin and meaning.

In our system, the presence of such a 
web pattern might have important consequences for stability and resilience. 
Species removal  (Brown and Heske, 1990; Brown, 1994) or the 
introduction of sharp shocks in population abundances due to transient 
perturbations might be strongly damped through the rapid transfer 
of information characteristic of these webs (Watts and Strogatz, 1998). 
In this sense, classic studies on ecosystem resilience and resistance 
might benefit from considering these topological constraints. Most 
studies suggest that the characteristic time of recovery from short, 
sharp shocks (as it would be the case for a pulse of nutrients) depends 
on the lenght of the food chain. Some enlightening theoretical studies 
based on the analysis of small-species models show this trend (Pimm and 
Lawton, 1977) where some simple topological arrangements between species 
are predefined. But perhaps some other regularities will emerge from our 
results when rich-species models are taken into account. 

A different question emerging from our study is how such food webs 
are constructed through community assembly rules (Drake, 1990a, 1990b). 
Since membership in a community 
is usually influenced by interactions among species, the dynamical process 
leading to a SW pattern of connections might eventually allow to understand 
how our reported results are explained. 
A possible dynamical process underlying the presence of scaling in 
food webs is preferential attachement (Barab\'asi and Albert, 1999).
This process of assembly consists in the addittion of new nodes which are
preferentially linked to already highly-connected nodes. 
This process has been argued
as the explanatory mechanism of the organization of different 
complex networks, including technological ones or the large
set of metabolic networks analysed by Jeong and collaborators (2000). But for
ecological networks (represented as food webs) is far from trivial to 
decide if preferential attachement is an appropriate, general assembly rule. 
The ubiquity of assembly rules has been recently discussed (Weiher and 
Keddy 1999) almost each one corresponding to each of the different
ecological communities studied. 

Even if a community is assembled 
by random addition of species, interactions can lead to non-random webs. 
If network homeostasis is a global constraint to species richness,
then a SW pattern might be an 'attractor' in community assembly
(Weiher and Keddy, 1999). Future models will require to
incorporate weighted, directed links and the scaling behavior of these 
graphs can have some important effects in extinction dynamics (Sol\'e and 
Montoya, in preparation) and might lead to some unexpected results. 
The SW topology might be related to underlying community
level constraints and might help understanding how biodiversity
emerges and persists.

\vspace{0.6 cm}

{\Large Acknowledgments}

\vspace{0.2 cm}

The authors thank Javier Gamarra and David Alonso for help in the analysis 
of some of the food webs and useful ``night crew'' discussions on theoretical 
ecology. Special thanks to Ramon Ferrer for his help with the graph analysis. 
We also thank Miguel Angel Rodriguez for useful talks and a careful reading
of the manuscript. This work has been supported by a grant
CICYT PB97-0693 and The Santa Fe Institute (RVS).

\vspace{0.5 cm}

\begin{enumerate}

\item
Adamic L. A. (1999) {\em The small world web}. Proceedings of the 
ECDL'99 Conference, pp.443-452 (Springer, Berlin).

\item
Albert, R., Jeong, H. and Barabasi, A-L (2000) Error and attack tolerance of
complex networks. Nature 406, 378-382.
Jeong, H., Tombor, B., Albert, R., Oltvai, Z.N. and Barabasi, A-L. (2000) The
large scale organization of metabolic networks. Nature {\bf 407}, 651-654.

\item
Amaral, L. A. N., Scala, A., Barth\'elemy, M. and Stanley, H. E. (2000) 
Classes of behavior of small-world networks. 
Proc. Nat. Acad. Sc. USA {\bf 97}, 11149-11152.

\item
Barab\'asi, L. A. and Albert, R. (1999) Emergence of Scaling in 
Random Networks, {\em Science}, {\bf 286}, 509-512

\item
Bollob\'as, B. (1985) {\em Random Graphs}. (Academic Press, London) 

\item 
Brown, J. H. and Heske, E. J. (1990) Control of a desert-grassland 
transition by a keystone rodent guild. {\em Science}, {\bf 250}, 1705-1707

\item 
Brown, J. H. (1994) Complex ecological systems, in {\it Complexity: Metaphors,
Models and Reality} Cowan, G., Pines, D., and Meltzer, D.,
eds., pp. 419-449, Addison Wesley.

\item
Cohen, J. E., Briand, F. and Newman, C. M. (1990) {\em Community 
Food Webs: Data and Theory}. Biomathematics, vol. 20 (Springer, Berlin)

\item
Drake, J. A. (1990a) The mechanics of community assembly rules. 
{\em J. Theor. Biol.}, {\bf 147}, 213-233.

\item
Drake, J. A. (1990b) Communities as assembled structures: do rules
govern pattern?. {\em Trends Ecol. Evol.}, {\bf 5}, 159-163.

\item
Hall, S. J. and Raffaelli, D. (1991) Food web patterns: lessons from a 
species-rich web.  {\em J. Anim. Ecol. }, {\bf 60}, 823-842

\item
Hanski, I. (1999) {\em Metapopulation Ecology}, Oxford U. Press.

\item
Huxham, M., Beaney, S. and Raffaelli, D. (1996) Do parasites reduce 
the chances of triangulation in real food webs? {\em OIKOS} {\bf 76}, 
284-300

\item
Martinez, N. D. (1991a) Constant connectance in community food webs. 
{\em Am. Nat.} {\bf 139}, 1208-121. 

\item
Martinez, N. D. (1991b) Artifacts or attributes? Effects of resolution on the 
Little Rock lake food web. {\em Ecol. Monographs} {\bf 61}, 367-392. 

\item 
May, R.M. (1974) {\it Stability and complexity in model ecosystems}. 
Princeton U. Press.

\item
Memmott, J., Martinez, N. D. and Cohen, J. E. (2000) Predators, parasitoids 
and pathogens: species richness, trophic generality and body sizes in a 
natural food web. {\em J. Anim. Ecol.} {\bf 69}, 1-15

\item
Pimm, S. L. and Lawton, J. H. (1977) The number of trophic levels 
in ecological communities. {\em Nature (Lond.)} {\bf 268}, 329-331  

\item
Pimm, S. L. and Lawton, J. H. (1980) Are food webs divided into 
compartments?  {\em J. Anim. Ecol. }, {\bf 49}, 879-898

\item
Pimm, S. L. (1982) {\em Ecological Food Webs}, Chapman and Hall. 

\item
Pimm, S. L., Lawton, J. H. and Cohen, J. E. (1991) Food web patterns and their 
consequences. {\em Nature (Lond.)} {\bf 350}, 669-674. 

\item
Pimm, S. L. (1991) {\it The Balance of Nature}. Chicago Press.

\item
Polis, G.A. and Strong, D.R. (1996) Food web complexity
and community dynamics. American Naturalist 147, 813-846.

\item
Sol\'e, R. V. et al. (1996) Phase Transitions and Complex Systems, 
{\em Complexity} {\bf 1} (4), 13-18

\item
Sol\'e, R. V. and Bascompte, J. (2001) {\em Complexity and Self-organization 
in Evolutionary Ecology}, Monographs in Population Biology.
(Princeton U. Press) (to appear)

\item
Solow, A. R. and Beet, A. R. (1998) On lumping species in food webs. 
{\em Ecology} {\bf 79}, 2013-2018 

\item
Solow, A. R., Costello, C. and Beet, A. R. (1999) On an early result 
on stability and complexity. {\em Am. Nat.} {\bf 154}, 587-588

\item 
Wagner, A. and Fell, D. (2000) The small world inside large metabolic 
networks, {\em Santa Fe Institute Working Paper} 00-07-041. 

\item
Warren, P. H. (1994) Making connections in food webs. 
{\em Trends Ecol. Evol.} {\bf 4}, 136-140 

\item
Watts, D. J. (1999) {\em Small Worlds}. Princeton U. Press

\item
Watts, D. J. and Strogatz, S. H. (1998) Collective Dynamics 
in 'small-world' networks, {\em Nature (Lond.)} {\bf 393}, 440-442

\item
Weiher, E. and Keddy, P., eds. (1999) {\em Ecological Assembly Rules}. 
Cambridge U. Press

\item
Winemiller, K.O. (1990) Spatial and temporal variation in 
tropical fish trophic networks. Ecol. Monog. {\ bf60}, 331-367.

\item
Williams, R. J. and Martinez, N. D. (2000) Simple rules yield complex 
food webs.  {\em Nature (Lond.)} {\bf 404}, 180-183

\item
Wilson, J.W. (1999) Guilds, functional types and ecological 
groups. Oikos 86, 507-522.

\end{enumerate}

\end{multicols}

\end{document}